\documentclass[aps,pra,showpacs,superscriptaddress,preprint]{revtex4}
\usepackage{amssymb}
\usepackage{amsmath}
\usepackage{graphicx}

\catcode`ð=\active
 \defð{\u{g}}
 \catcode`Ð=\active
 \defÐ{\u{G}}
 \catcode`Ý=\active
\defÝ{\. I}
 \catcode`ö=\active
\defö{\"{o}}
 \catcode`Ö=\active
 \defÖ{\"O}
 \catcode`ü=\active
 \defü{\"{u}}
 \catcode`Ü=\active
 \defÜ{\"{U}}
 \catcode`Þ=\active
\defÞ{\c{S}}
 \catcode`þ=\active
 \defþ{\c{s}}
 \catcode`ý=\active
 \defý{{\i}}
 \catcode`ç=\active
\defç{\d{c}}
 \catcode`Ç=\active
\defÇ{\d{C}}

\input{tcilatex}

\begin{document}

\title{Rotation and vibration of diatomic molecule in the
spatially-dependent mass Schr\"{o}dinger equation with generalized $q$%
-deformed Morse potential }
\author{Sameer M. Ikhdair}
\email[E-mail: ]{sikhdair@neu.edu.tr; sameer@neu.edu.tr}
\affiliation{Physics Department, Near East University, Nicosia, Cyprus, Turkey}
\date{%
\today%
}

\begin{abstract}
The analytic solutions of the spatially-dependent mass Schr\"{o}dinger
equation of diatomic molecules with the centrifugal term $l(l+1)/r^{2}$ for
the generalized $q$-deformed Morse potential are obtained approximately by
means of a parametric generalization of the Nikiforov-Uvarov (NU) method
combined with the Pekeris approximation scheme. The energy eigenvalues and
the corresponding normalized radial wave functions are calculated in closed
form with a physically motivated choice of a reciprocal Morse-like mass
function, $m(r)=m_{0}/\left( 1-\delta e^{-a(r-r_{e})}\right) ^{2},$ $0\leq
\delta <1,$ where $a$ and $r_{e}$ are the range of the potential and the
equilibrium position of the nuclei$.$ The constant mass case when $\delta
\rightarrow 0$ is also studied. The energy states for $H_{2},$ $LiH,$ $HCl$
and $CO$ diatomic molecules are calculated and compared favourably well with
those obtained by using other approximation methods for arbitrary
vibrational $n$ and rotational $l$ quantum numbers.

\end{abstract}

\maketitle

\newpage

\section{Introduction}

It is well known that a reasonable potential to describe the molecular
vibrational spectra of diatomic and even polyatomic molecules successfully
is the Morse potential [1-3]. There has been considerable current interest
in the use and application of the Morse oscillator model to the
investigation of local vibrational modes, with a particular emphasis on
those highly excited vibrational levels which are accessible by laser
spectroscopy [2]. An effective potential, which is the sum of the
centrifugal potential term that depends on the angular momentum $l$ and the
Morse potential, has been used as a model for such interactions. It is
referred to as the rotational potential. The radial non-relativistic and
relativistic wave equations with this potential are exactly solvable for $%
l=0 $ case [1]. However, for $l\neq 0$ analytic exact solution cannot be
determined, only numerical solutions are possible where several
approximation techniques have been proposed and extensively used with
varying degrees of accuracy and stability [4-17]. The most widely known
approximation was devised by Pekeris [5] and used to obtain the
semiclassical solutions [6-17]. This approximation [10,13] is based on the
expansion of the centrifugal part in exponential terms with exponents that
depend on an inter-nuclear distance parameter. This is why the Pekeris
approximation is valid only for very small spatial variations from the
inter-nuclear separation (i.e., for lower vibrational and rotational
energy). Other methods that have also been used include the variational (V)
method with the Pekeris approximation [6], supersymmetry (SUSY) using the
Pekeris approximation [7], the hypervirial (HV) perturbation method with the
effective potential and without the Pekeris approximation [8], the shifted $%
1/N$ expansion (SE) [9] and the modified shifted large $1/N$ approach (MSE)
[9] which is very cumbersome to solve because there are several complex
coefficients. There are needs of analytic solutions for these levels in
order to be used for static and other applications sometimes using
derivatives of this energy level functions. Other methods, which are
semi-analytic, have also been used. These include, the Nikiforov-Uvarov (NU)
[10] and the asymptotic iteration method (AIM) [12], using the Pekeris
approximation, where the energy eigenvalues are obtained by simple
transformation of the wave equation and the wave function is calculated
iteratively. The exact quantization rule (EQR) [13,14] and the tridiagonal
J--matrix representation (TJM) [15] which split original Hamiltonian into
two parts as $H=H_{0}+V$ where $H_{0}$ is the part of the Hamiltonian that
could be treated analytically while the remaining part, $V,$ has to be
treated numerically. Recent techniques of approximation denoted as two-point
quasi-rational approximation technique (TQA) [16], which is considered as an
extension of the Pad\'{e} procedure, have been used in energy hydrogenic's
levels determination of Zeeman effects and Coulomb potential with harmonics
and quadratic terms, was used for the hydrogen molecule, the results of Duff
and Rabitz (DR) [17] and the Fourier grid Hamiltonian method (FGH) [18], etc.

On the other hand, the studies of the non-relativistic and relativistic wave
equations with position-dependent mass (PDM) for central physical potentials
have attracted much attentions [18-29]. Such solutions are very useful in
different fields of material science and condensed matter physics, such as
semiconductors [30], quantum well and quantum dots [31], $^{3}He$ clusters
[32], quantum liquids [33], graded alloys and semiconductor heterostructures
[34], etc.

The NU method [35] and other methods have also been used to solve the Schr%
\"{o}dinger equation [27,28], relativistic $D$-dimensional Klein-Gordon
equation [29,36] and Dirac equation [37] with PDM case.

In the last few years, a considerable interest in the use and application of
the mass dependence on the inter-nuclear distance in solving the
non-relativistic and relativistic wave equations with various central
potentials including the Morse potential have been revived. It provides us a
reasonable first approximation for vibrational dynamics of diatomic
molecules including the effects of anharmonicity and bond dissociation, it's
incorrect behavior at large interatomic separations results in a relatively
poor description of highly-excited vibrational levels, which are of much
concern particularly for laser spectroscopy [2]. In other context, local
model descriptions of molecular vibrations are required to understand the
behavior of molecules for high overtone and combination levels [2,38]. The
development of new experimental methods has unveiled a whole new range of
molecular phenomena, including localization, intra-molecular energy
redistribution and isomerization [39,40]. Even highly symmetrical molecules
can develop localization at higher energy [2], a feature difficult to
understand from the point of view of the traditional normal mode models of
constant mass. In their pioneering work, Child \textit{et al} have
emphasized the need for introducing anharmonicity at the local level as a
more natural starting point [2].

Since the realistic diatomic potentials are more accurately modelled by the
perturbed or $q$-deformed Morse potentials, we believe that the present
approach will produce much more accurate information about the structure,
dynamics and even the issue of energy resonances while dealing with the
diatomic molecules. This is going to be the subject of investigation in the
current study. In order to make the treatment very simple, we set out to
employ the parametric generalization of the NU method combined with the
Pekeris approximation scheme to solve the PDM radial Scr\"{o}dinger
equation\ for a generalized $q$-deformed Morse potential with arbitrary
rotational angular quantum number. We propose and use a reciprocal
Morse-like effective mass distribution function physically derived from the
exact pseudo-spin symmetric solution of the Dirac equation [37].
Furthermore, the motivation for the choice is the nature of the field which
is supposed to prevail between the interacting nuclei. This suitable choice
may result in a solvable PDM Schr\"{o}dinger equation for it's energy
eigenvalues and their corresponding wave functions. Besides, it allows one
to get a compact analytical expression and transforms the corresponding
Morse Hamiltonian into the constant-mass problem.

The present paper is organized as follows. In sect. 2, we briefly outline
the basics of the NU method and supplement the parametric generalization of
the method to exponential-type potentials in Appendix A. In sect. 3, we
calculate the approximate analytic NU bound state energy eigenvalues and
normalized wave functions of the PDM Schr\"{o}dinger equation with the
generalized $q$-deformed Morse potential for any $l$-state. The Pekeris
analytic expansions are cited in Appendix B. In sect. 4, we calculate the
numerical energy states for the various vibrational $n,$ rotational $l$
quantum numbers and $q$ deformation parameter for four different diatomic
molecules $CO,$ $LiH,$ $H_{2}$ and $HCl$ in the constant mass limit ($\delta
\rightarrow 0).$ Finally, sect. 5 contains the relevant conclusions.

\section{NU Method}

The NU method has been used to solve the Schr\"{o}dinger [28], Klein-Gordon
[29] and Dirac [37] wave equations for central and non-central potentials.
Let us briefly outline the basic concepts of the method [35]. This method
was proposed to solve the second-order linear differential equation of the
hypergeometric-type: 
\begin{equation}
\sigma ^{2}(z)u^{\prime \prime }(z)+\sigma (z)\widetilde{\tau }(z)u^{\prime
}(z)+\widetilde{\sigma }(z)u(z)=0,
\end{equation}%
where the prime denotes the differentiation with respect to $z,$ $\sigma (z)$
and $\widetilde{\sigma }(z)$ are analytic polynomials, at most
second-degree, and $\widetilde{\tau }(s)$ is a first-degree polynomial$.$
Let us discuss the exact particular solution of Eq. (1) by choosing%
\begin{equation}
u(z)=y_{n}(z)\phi _{n}(z),
\end{equation}%
resulting in a hypergeometric type equation of the form: 
\begin{equation}
\sigma (z)y_{n}^{\prime \prime }(z)+\tau (z)y_{n}^{\prime }(z)+\lambda
y_{n}(z)=0.
\end{equation}%
The first part $y_{n}(z)$ is the hypergeometric-type function whose
polynomial solutions are given by the Rodrigues relation%
\begin{equation}
y_{n}(z)=\frac{A_{n}}{\rho (z)}\frac{d^{n}}{dz^{n}}\left[ \sigma ^{n}(z)\rho
(z)\right] ,
\end{equation}%
where $A_{n}$ is a normalization factor and $\rho (z)$ is the weight
function satisfying the condition%
\begin{equation}
\left[ \sigma (z)\rho (z)\right] ^{\prime }=\tau (z)\rho (z),
\end{equation}%
with 
\begin{equation}
\tau (z)=\widetilde{\tau }(z)+2\pi (z),\tau ^{\prime }(z)<0.
\end{equation}%
Since $\rho (z)>0$ and $\sigma (z)>0,$ the derivative of $\tau (z)$ has to
be negative [35] which is the main essential condition for any choice of
particular bound-state solutions. The other part of the wave function is
defined as a logarithmic derivative:%
\begin{equation}
\frac{\phi ^{\prime }(z)}{\phi (z)}=\frac{\pi (z)}{\sigma (z)},
\end{equation}%
where%
\begin{equation}
\pi (z)=\frac{1}{2}\left[ \sigma ^{\prime }(z)-\widetilde{\tau }(z)\right]
\pm \sqrt{\frac{1}{4}\left[ \sigma ^{\prime }(z)-\widetilde{\tau }(z)\right]
^{2}-\widetilde{\sigma }(z)+k\sigma (z)},
\end{equation}%
with%
\begin{equation}
k=\lambda -\pi ^{\prime }(z).
\end{equation}%
The determination of $k$ is the key point in the calculation of $\pi (z),$
for which the discriminant of the square root in the last equation is set to
zero. This results in the polynomial $\pi (z)$ which is dependent on the
transformation function $z(r).$ Also, the parameter $\lambda $ defined in
Eq. (9) takes the form%
\begin{equation}
\lambda =\lambda _{n}=-n\tau ^{\prime }(z)-\frac{1}{2}n\left( n-1\right)
\sigma ^{\prime \prime }(z),\ \ \ n=0,1,2,\cdots .
\end{equation}%
At the end, the energy equation and consequently it's eigenvalues can be
obtained by comparing Eqs. (9) and (10).

Let us now construct a parametric generalization of the NU method valid for
any central and non-central exponential-type potential. Comparing the
following generalized hypergeometric equation 
\begin{equation}
\left[ z\left( 1-c_{3}z\right) \right] ^{2}u^{\prime \prime }(z)+\left[
z\left( 1-c_{3}z\right) \left( c_{1}-c_{2}z\right) \right] u^{\prime
}(z)+\left( -Az^{2}+Bz-C\right) u(z)=0,
\end{equation}%
with Eq. (1), we obtain%
\begin{equation}
\widetilde{\tau }(z)=c_{1}-c_{2}z,\text{ }\sigma (z)=z\left( 1-c_{3}z\right)
,\text{ }\widetilde{\sigma }(z)=-Az^{2}+Bz-C,
\end{equation}%
where the parameters $c_{1},$ $c_{2},$ $c_{3},$ $A,$ $B$ and $C$ are
constants to be determined during the solution later. Thus, by following the
method, we may obtain all the analytic polynomials and their relevant
constants necessary for the solution of a radial wave equation. These
analytic expressions are cited in Appendix A.

\section{NU Solutions for the Generalized $q$-Deformed Morse Potential}

Choosing the separated atoms limit as the zero of energy, the generalized $q$%
-deformed Morse potential has the following form [1,41,42]:%
\begin{equation}
V_{M}(r)=D_{e}\left[ q-\exp (-\alpha x)\right] ^{2}=V_{1}\exp (-2\alpha
x)-V_{2}\exp (-\alpha x)+V_{3},
\end{equation}%
where $\alpha =ar_{e},$ $x=(r-r_{e})/r_{e},$ $V_{1}=D_{e},$ $V_{2}=2qD_{e}$
and $V_{3}=q^{2}D_{e}$ where subscripts $1$ and $2$ refer to the repulsive
and attractive terms, respectively$.$ The range of the deformation parameter 
$q$ in the above potential was taken as $q>0$ by Ref. [43] and has been
extended to $-1\leq q<0$ or $q>0$\ or even complex by Ref. [44]. The above
potential contains three adjustable positive parameters; the parameter $%
D_{e} $ corresponds to the depth of the potential well, $a$ is related with
the range of the potential and $r_{e}$ is the equilibrium position of the
nuclei. At $r=r_{e},$ it has a minimum value at $V_{M}(r_{e})=D_{e}(q-1)^{2}$
and approaches $qD_{e}$ exponentially for large $r.$ If $1/\alpha $ is
somewhat smaller than the equilibrium distance $r_{e},$ it becomes large
(but not infinite) as $r\rightarrow 0.$ Here, we refer to the above
potential as that applicable to the diatomic molecules. For the case of
multiatomic molecules, the generalization of Morse potentials can be
expressed by the Empirical Valence Bond (EVB) approach [45].

We now study the Schr\"{o}dinger equation with the PDM for the potential
form (13), which can be expressed as [20,26,37]

\begin{equation}
\overrightarrow{\nabla }\left( \frac{1}{m}\overrightarrow{\nabla }\psi
(r)\right) +2\left[ E_{nl}-V(r)\right] \psi (\mathbf{r})=0,
\end{equation}%
where $m=m(r)$ and $E_{nl}$ are the real mass function and the energy
eigenvalues, respectively. For spherical symmetry, the wave function could
be separated to the following form [46-49] 
\begin{equation}
\psi (\mathbf{r})=\frac{1}{r}R_{l}(r)Y_{lm}(\theta ,\phi ).
\end{equation}%
which gives the following radial wave equation:%
\begin{equation}
\left\{ \frac{d^{2}}{dr^{2}}+\frac{m^{\prime }}{m}\left( \frac{1}{r}-\frac{d%
}{dr}\right) -\frac{l(l+1)}{r^{2}}+\frac{2m}{\hbar ^{2}}\left[ E_{nl}-V(r)%
\right] \right\} R_{l}(r)=0,
\end{equation}%
where $m^{\prime }=dm(r)/dr$ and $R_{l}(r)$ are the radial wave functions$.$
It is worth noting that for $m^{\prime }=0$ case$,$ the above equation
reduces to the well-known equation with constant mass used in Refs. [49-51].
Furthermore, the above equation with the use of transformation%
\begin{equation}
R_{l}(r)=\sqrt{m}u_{l}(r),
\end{equation}%
reduces to the Schr\"{o}dinger-like equation:%
\begin{equation*}
-\frac{d^{2}u_{l}(r)}{dr^{2}}+V_{eff}(r)u_{l}(r)=\frac{2mE_{nl}}{\hbar ^{2}}%
u_{l}(r),\text{ }u_{l}(0)=0,
\end{equation*}%
\begin{equation}
V_{eff}(r)=-\frac{m^{\prime \prime }}{2m}+\frac{3}{4}\left( \frac{m^{\prime }%
}{m}\right) ^{2}-\frac{m^{\prime }}{m}\frac{1}{r}+\frac{l(l+1)}{r^{2}}+\frac{%
2m}{\hbar ^{2}}V(r),
\end{equation}%
where the interaction potential in the above equation is taken to be the
generalized $q$-deformed Morse potential (13). In addition, for the bound
state solutions, i.e., real energy eigenvalues$,$ the wave functions $%
u_{l}(r)$ need to be finite near $r=0$ and $r\rightarrow \infty .$

In order to find an approximation [4] for the centrifugal potential term $%
l(l+1)/r^{2}$ and $\ $for $1/r,$ we apply the Pekeris approximation [5,10].
We further make the appropriate parameter change $r=r_{e}(1+x)$ and perform
the expansion around $x=0$ ($r=r_{e}$) to second-order in $x$ ($r/r_{e}$) at
low-excitation energy (i.e., $r\approx r_{e}$). Following the approach in
Ref. [10], we make the convenient expansions and obtain expansion parameters 
$a_{i}$ and $b_{i}$ ($i=0,1,2)$ ) as demonstrated in Appendix B. Thus, with
the Pekeris approximations, the semi-analytic solution of the Schr\"{o}%
dinger equation for the generalized $q$-deformed Morse potential ($l\neq 0$)
has become amenable. For the present potential model, the exponential choice
for the PDM allows one to transform the corresponding Morse Hamiltonian into
the constant mass problem. Since the field between the two interacting
nuclei is a short-range field of which the Morse potential is an example, we
are tempted to set the PDM function as a reciprocal Morse-like ansatz having
the form [52,53]%
\begin{equation}
m=\frac{m_{0}}{\left( 1-\delta z\right) ^{2}},\text{ }m^{\prime }=-\frac{%
2m_{0}\delta \alpha z}{\left( 1-\delta z\right) ^{3}},\text{ }m^{\prime
\prime }=\frac{2m_{0}\delta \alpha ^{2}z}{\left( 1-\delta z\right) ^{3}}+%
\frac{6m_{0}\delta ^{2}\alpha ^{2}z^{2}}{\left( 1-\delta z\right) ^{4}},
\end{equation}%
where $z=\exp (-\alpha x)$ $\in $ $\left( 0,1\right) ,$ and $0\leq \delta <1$%
. The above mass function works well in the present model and it results in
a solvable wave equation. However, in this choice, there is no loss of
generality when the value of the parameter $\delta $ is taken to be small
(i.e., $\delta \rightarrow 0$). Then, Eq. (18) with the help of Eq. (19),
and the expansions made in Appendix B reduces to,%
\begin{equation*}
u_{l}^{\prime \prime }(z)+\frac{1}{z}u_{l}^{\prime }(z)+\frac{1}{z^{2}\left(
1-\delta z\right) ^{2}}\left\{ \frac{2m_{0}}{\hbar ^{2}a^{2}}\left(
E_{nl}-V_{3}\right) -\frac{\gamma a_{0}}{a^{2}}+\left( \frac{2m_{0}}{\hbar
^{2}a^{2}}V_{2}-\frac{\gamma a_{1}}{a^{2}}+S\delta \right) z\right.
\end{equation*}%
\begin{equation}
-\left. \left( \frac{2m_{0}}{\hbar ^{2}a^{2}}V_{1}+\frac{\gamma a_{2}}{a^{2}}%
+P\delta +Q\delta ^{2}\right) z^{2}\right\} u_{l}(z)=0,
\end{equation}%
with%
\begin{equation}
S=r_{e}^{2}-\frac{2b_{0}}{a}+\frac{2\gamma a_{0}}{a^{2}},\text{ }P=\frac{%
2b_{1}}{a}-\frac{2\gamma a_{1}}{a^{2}},\text{ }Q=r_{e}^{2}-\frac{2b_{0}}{a}+%
\frac{\gamma a_{0}}{a^{2}},
\end{equation}%
where $a_{i}$ and $b_{i}$ ($i=0,1,2$) are given in Appendix B and $%
u_{l}(z)=u_{l}(r).$ Defining the following parameters 
\begin{equation*}
\varepsilon _{nl}=\frac{1}{a}\sqrt{\gamma a_{0}+\frac{2m_{0}}{\hbar ^{2}}%
\left( V_{3}-E_{nl}\right) },\text{ \ }\beta _{1}=\frac{1}{a^{2}}\left( 
\frac{2m_{0}}{\hbar ^{2}}V_{1}+\gamma a_{2}\right) +P\delta +Q\delta ^{2},
\end{equation*}%
\begin{equation}
\beta _{2}=\frac{1}{a^{2}}\left( \frac{2m_{0}}{\hbar ^{2}}V_{2}-\gamma
a_{1}\right) +S\delta ,
\end{equation}%
reduces Eq. (20) into the form: 
\begin{equation}
\frac{d^{2}u_{l}(z)}{dz^{2}}+\frac{1}{z}\frac{du_{l}(z)}{dz}+\frac{1}{\left[
z(1-\delta z)\right] ^{2}}\left( -\beta _{1}z^{2}+\beta _{2}z-\varepsilon
_{nl}^{2}\right) u_{l}(z)=0.
\end{equation}%
In comparing Eq. (23) with Eq. (1), we follow Appendix A to obtain the
specific values of the parametric constants and hence present them in Table
1 for the currently used potential model. Also, with the aid of Table 1, the
key polynomials given in Appendix A now take the following particular
analytic forms: 
\begin{equation}
\pi (z)=\varepsilon _{nl}-\frac{\delta }{2}\left( 1+2\varepsilon _{nl}+\xi
\right) z,
\end{equation}%
\begin{equation}
k=\beta _{2}-\delta \left( 2\varepsilon _{nl}+\xi \right) \varepsilon _{nl},
\end{equation}%
\begin{equation}
\tau (z)=1+2\varepsilon _{nl}-\delta \left( 2+2\varepsilon _{nl}+\xi \right)
z,
\end{equation}%
with $\tau ^{\prime }(z)=-\delta \left( 2+2\varepsilon _{nl}+\xi \right) <0,$
where $\xi =\sqrt{1+4\varepsilon _{nl}^{2}+\frac{4}{\delta }\left( \frac{%
\beta _{1}}{\delta }-\beta _{2}\right) }.$ We insert the values of the
constants given in Table 1 into the energy equation cited in Appendix A to
obtain%
\begin{equation}
\varepsilon _{nl}=\frac{1}{2}\frac{n(n+1)\delta -2\left( n+\frac{1}{2}%
\right) \sqrt{\beta _{1}}+\beta _{2}}{\sqrt{\beta _{1}}-\left( n+\frac{1}{2}%
\right) \delta }.
\end{equation}%
The above equation can be written more explicitly for the rotational
bound-state energy eigenvalues as%
\begin{equation}
E_{nl}=V_{3}+\frac{\hbar ^{2}l(l+1)a_{0}}{2m_{0}r_{e}^{2}}-\frac{\hbar
^{2}a^{2}}{8m_{0}}\left[ \frac{n(n+1)\delta -2\left( n+\frac{1}{2}\right) 
\sqrt{\beta _{1}}+\beta _{2}}{\sqrt{\beta _{1}}-\left( n+\frac{1}{2}\right)
\delta }\right] ^{2},\text{ }n,l=0,1,2,\cdots ,
\end{equation}%
and after substituting the particular values of the parameters $\beta _{1}$
and $\beta _{2}$ from Eq. (22), we finally obtain the energy eigenvalues
expressed in terms of the strength parameters $V_{i}$ ($i=1,2,3$) as%
\begin{equation*}
E_{nl}=V_{3}+\frac{\hbar ^{2}l(l+1)a_{0}}{2m_{0}r_{e}^{2}}-\frac{\hbar
^{2}a^{2}}{8m_{0}}\text{ }
\end{equation*}%
\begin{equation}
\times \left[ \frac{n(n+1)\delta -2\left( n+\frac{1}{2}\right) \sqrt{\frac{1%
}{a^{2}}\left( \frac{2m_{0}}{\hbar ^{2}}V_{1}+\frac{l(l+1)a_{2}}{r_{e}^{2}}%
\right) +P\delta +Q\delta ^{2}}+\frac{1}{a^{2}}\left( \frac{2m_{0}}{\hbar
^{2}}V_{2}-\frac{l(l+1)a_{1}}{r_{e}^{2}}\right) +S\delta }{\sqrt{\frac{1}{%
a^{2}}\left( \frac{2m_{0}}{\hbar ^{2}}V_{1}+\frac{l(l+1)a_{2}}{r_{e}^{2}}%
\right) +P\delta +Q\delta ^{2}}-\left( n+\frac{1}{2}\right) \delta }\right]
^{2},
\end{equation}%
where $S,$ $P$ and $Q$ are constant parameters given in Eq. (21). In
particular, for the constant mass case, i.e., in the limit $\delta
\rightarrow 0$ ($m\rightarrow m_{0}$)$,$ we can easily reduce the above
equation to,%
\begin{equation}
\varepsilon _{nl}=\frac{\beta _{2}}{2\sqrt{\beta _{1}}}-\left( n+\frac{1}{2}%
\right) ,
\end{equation}%
or more explicitly as%
\begin{equation*}
E_{nl}=V_{3}+\frac{\hbar ^{2}}{2\mu r_{e}^{2}}l(l+1)\left( 1-\frac{3}{ar_{e}}%
+\frac{3}{a^{2}r_{e}^{2}}\right)
\end{equation*}%
\begin{equation}
-\frac{\hbar ^{2}a^{2}}{2\mu }\left[ \sqrt{\frac{2\mu }{\hbar ^{2}a^{2}}}%
\frac{\frac{V_{2}}{2}-\frac{\hbar ^{2}}{2\mu r_{e}^{2}}l(l+1)\left( \frac{2}{%
ar_{e}}-\frac{3}{a^{2}r_{e}^{2}}\right) }{\sqrt{V_{1}-\frac{\hbar ^{2}}{2\mu
r_{e}^{2}}l(l+1)\left( \frac{1}{ar_{e}}-\frac{3}{a^{2}r_{e}^{2}}\right) }}%
-\left( n+\frac{1}{2}\right) \right] ^{2},
\end{equation}%
with $\mu =m_{1}m_{2}/(m_{1}+m_{2})$ denotes the reduced mass for the
diatomic molecule. The above equation represents the energy eigenvalues for
the generalized $q$-deformed Morse potential [42]. The vibrational bound
state energy levels for $s$-waves $(l=0)$ are%
\begin{equation}
E_{n}=V_{3}-\frac{1}{4\kappa ^{2}}\left[ 1+2n-\eta \kappa \right] ^{2},
\end{equation}%
with%
\begin{equation}
\eta =\frac{V_{2}}{\sqrt{V_{1}}},\text{ }\kappa =\frac{r_{e}}{\alpha \hbar }%
\sqrt{2\mu },\text{ }n_{\max }\leq \frac{1}{2}\left( \frac{\sqrt{2\mu }}{%
\hbar a}\frac{V_{2}}{\sqrt{V_{1}}}-1\right) ,
\end{equation}%
where $V_{i}$ ($i=1,2,3$) are defined after Eq. (13). Therefore, the
vibrational quantum number $n=0,1,2,\cdots ,n_{\max },$ where $n_{\max }$ is
the number of bound states for the whole bound spectrum near the continuous
zone. Thus, $n_{\max }$ cannot be infinite, which is reflected in the above
condition. Therefore, the Morse eigenfunctions for real systems do not form
an infinite-dimensional Lie algebra.

Let us now turn to the calculations of the corresponding wave functions in
the varying mass case. Referring to Appendix A and Table 1, we find the
explicit form of the weight function as 
\begin{equation}
\rho (z)=z^{2\varepsilon _{nl}}(1-\delta z)^{\xi },
\end{equation}%
where $\varepsilon _{nl}$ is given in Eq. (27). The above weight function
gives the first part of the wave functions:%
\begin{equation}
y_{n}(z)\rightarrow P_{n}^{(2\varepsilon _{nl},\xi )}(1-2\delta z),
\end{equation}%
and hence the second part of the wave functions can be found as%
\begin{equation}
\phi (z)\rightarrow z^{\varepsilon _{nl}}(1-\delta z)^{\frac{1}{2}+\frac{_{1}%
}{2}\xi }.
\end{equation}%
Hence, the unnormalized wave functions are being expressed in terms of the
Jacobi polynomials as%
\begin{equation}
u_{l}(z)=\mathcal{N}_{n}z^{\varepsilon _{nl}}(1-\delta z)^{\frac{1}{2}\left(
1+\xi \right) }P_{n}^{(2\varepsilon _{nl},\xi )}(1-2\delta z),\text{ }0<%
\text{ }\delta <1,
\end{equation}%
where the normalization constant is being calculated using the formulas
placed in Appendix A as%
\begin{equation*}
\mathcal{N}_{n}=\left[ \frac{\Gamma \left( 2\varepsilon _{nl}+1\right)
\Gamma \left( \xi +2\right) }{\alpha \delta ^{\varepsilon _{nl}}\Gamma
\left( n\right) }\dsum\limits_{p=0}^{\infty }\frac{(-1)^{p}\Gamma
(n+p)\left( n+1+2\varepsilon _{nl}+\xi \right) _{p}}{p!(p+2\varepsilon
_{nl})\Gamma (p+2\varepsilon _{nl}+\xi +2)}\right.
\end{equation*}%
\begin{equation}
\times \left. 
\begin{array}{c}
_{3}F_{2}%
\end{array}%
\left( p+2\varepsilon _{nl},-n,n+2\varepsilon _{nl}+\xi +1;p+2\varepsilon
_{nl}+\xi +2;1+2\varepsilon _{nl};1\right) \right] ^{-1/2},
\end{equation}%
with $\varepsilon _{nl}$ and $\xi $ defined in Eq. (27) and after Eq. (26),
respectively. $\left( x\right) _{p}$ is the Pochhammer symbols defined as%
\begin{equation}
\left( x\right) _{p}=\frac{\Gamma \left( x+p\right) }{\Gamma \left( x\right) 
}.
\end{equation}%
Thus, the total radial part of the wave functions of the generalized $q$%
-deformed potential is 
\begin{equation*}
\psi _{l}(r)=\mathcal{N}_{n}\frac{1}{r}\left[ \exp \left[ -a(r-r_{e})\right] %
\right] ^{\varepsilon _{nl}}(1-\delta \exp \left[ -a(r-r_{e})\right] )^{-%
\frac{1}{2}+\frac{_{1}}{2}\xi }
\end{equation*}%
\begin{equation}
\times P_{n}^{(2\varepsilon _{nl},\xi )}(1-2\delta \exp \left[ -a(r-r_{e})%
\right] ),\text{ }0<\text{ }\delta <1,
\end{equation}%
where $\mathcal{N}_{n}$ is defined in Eq. (38).

The constant mass case should be treated separately. To avoid repetition, we
can use our previous calculations to find an explicit form for the weight
function as [50,51]%
\begin{equation}
\rho (z)=z^{2\varepsilon _{nl}}\exp \left[ -2\sqrt{\beta _{1}}z\right] ,
\end{equation}%
which gives the Laguerre polynomials:%
\begin{equation}
y_{n}(z)\rightarrow z^{-2\varepsilon _{nl}}\exp \left[ 2\sqrt{\beta _{1}}z%
\right] \frac{d^{n}}{dz^{n}}(z^{n+2\varepsilon _{nl}}\exp \left[ -2\sqrt{%
\beta _{1}}z\right] )\rightarrow L_{n}^{2\varepsilon _{nl}}(y),
\end{equation}%
where $y=2\sqrt{\beta _{1}}z.$ The second part of the wave functions can be
found as%
\begin{equation}
\phi (z)\rightarrow z^{\varepsilon _{nl}}\exp \left[ -\sqrt{\beta _{1}}z%
\right] ,
\end{equation}%
Hence, the un-normalized wave functions expressed in terms of the Laguerre
polynomials read%
\begin{equation}
R_{l}(r)=N_{n}\left( 2\sqrt{\beta _{1}}\right) ^{-\varepsilon
_{nl}}y^{\varepsilon _{nl}}\exp \left( -\frac{y}{2}\right)
L_{n}^{2\varepsilon _{nl}}(y),
\end{equation}%
where $\varepsilon _{nl}$ is given in Eq. (27) and $y=2\sqrt{\beta _{1}}\exp %
\left[ -a(r-r_{e})\right] $ [10]$.$

To demonstrate the importance of adjusting the three potential strength
parameters $V_{i}$ ($i=1,2,3$)\ and $\alpha $ for real and/or complex values
in any possible numerical calculation, we now discuss four special cases of
the Morse potential given in (13) which are of much concern to the readers
[41,54,55].

\subsection{Generalized Vibrational Morse potential}

We consider the generalized vibrational Morse potential defined by [41,54,55]%
\begin{equation}
V_{M}(x)=V_{1}e^{-2\alpha x}-V_{2}e^{-\alpha x}+V_{3},\text{ }V_{1}=D,\text{ 
}V_{2}=2qD,\text{ }V_{3}=0,
\end{equation}%
and find the vibrational bound state energy spectrum as%
\begin{equation}
E_{n}=-\alpha ^{2}E_{0}\left[ \lambda q-n-\frac{1}{2}\right] ^{2},\text{ }%
n=0,1,2,\cdots ,n_{\max }
\end{equation}%
\begin{equation}
\lambda ^{2}=\frac{D}{\alpha ^{2}E_{0}},\text{ }n_{\max }\leq \frac{1}{2}%
\left( 2\lambda q-1\right) ,
\end{equation}%
where the derived quantity $E_{0}=\frac{\hbar ^{2}}{2\mu r_{e}^{2}}$ $(eV)$
with the following condition on the deformation parameter $q=$ $\frac{1}{%
2\lambda }$ for the final vibrational bound-state. The above result is
identical to Eq. (30) of Ref. [55]. Also, the wave functions are calculated
as%
\begin{equation}
R_{n}(x)=A_{n}\exp \left\{ -\alpha \left( \lambda q-n-\frac{1}{2}\right)
x-\lambda e^{-\alpha x}\right\} L_{n}^{2\left( \lambda q-n-\frac{1}{2}%
\right) }\left( 2\lambda e^{-\alpha x}\right) .
\end{equation}%
where $A_{n}$ is a normalizing factor.

\subsection{Non-PT Symmetric and Non-Hermitian Morse Case}

Following Refs. [54,55], let us assume the potential strength parameters $%
V_{1}=(A_{1}+iB_{1})^{2},$ $V_{2}=(2C_{1}+1)(A_{1}+iB_{1}),$ $V_{3}=0$ and $%
\alpha =1$ where $A_{1},$ $B_{1}$ and $C_{1}$ are real constants and $i=%
\sqrt{-1}.$ Under appropriate changes of parameters, the potential (13)
turns to become a complex Morse-like potential: 
\begin{equation}
V(x)=-D\left[ e^{-2x}+i\widehat{D}e^{-x}\right] .
\end{equation}%
Hence, we can get the vibrational real bound state energy spectrum given by%
\begin{equation}
E_{n}=-E_{0}\left[ \frac{1}{2}\widehat{D}\kappa _{1}-n-\frac{1}{2}\right]
^{2},\text{ }n=0,1,2,\cdots ,n_{\max },
\end{equation}%
with%
\begin{equation}
\text{ }\kappa _{1}=\frac{r_{e}}{\hbar }\sqrt{2\mu D},\text{ }n_{\max }\leq 
\frac{1}{2}\left( \frac{r_{e}}{\hbar }\widehat{D}\sqrt{2\mu D}-1\right) ,
\end{equation}%
which is similar to Eq. (41) of Ref. [55] and the wave functions as%
\begin{equation}
R_{n}(x)=B_{n}(2\kappa _{1})^{-\left( \frac{1}{2}\widehat{D}\kappa _{1}-%
\frac{1}{2}-n\right) }\left( 2\kappa _{1}e^{-x}\right) ^{\left( \frac{1}{2}%
\widehat{D}\kappa _{1}-\frac{1}{2}-n\right) }e^{-\kappa _{1}\exp
(-x)}L_{n}^{2\left( \frac{1}{2}\widehat{D}\kappa _{1}-\frac{1}{2}-n\right)
}\left( 2\kappa _{1}e^{-x}\right) ,
\end{equation}%
where $B_{n}$ is a normalizing factor.

\subsection{The First Type of PT-Symmetric and Non-Hermitian Morse case}

We consider the same strength parameters as in the previous case but $\alpha
=i,$ the potential (43) becomes%
\begin{equation}
V(x)=-D\left[ e^{-2ix}+i\widehat{D}e^{-ix}\right] .
\end{equation}%
Following the same procedure as before, we get no real spectrum for this
kind of potentials:%
\begin{equation}
E_{n}=E_{0}\left[ \frac{1}{2}\widehat{D}\kappa _{2}-n-\frac{1}{2}\right]
^{2},\text{ }n=0,1,2,\cdots ,n_{\max },
\end{equation}%
with%
\begin{equation}
\text{ }\kappa _{2}=\frac{r_{e}}{i\hbar }\sqrt{2\mu D},\text{ }n_{\max }\leq 
\frac{1}{2}\left( \frac{r_{e}}{i\hbar }\widehat{D}\sqrt{2\mu D}-1\right) .
\end{equation}%
Also, this gives the wave functions as%
\begin{equation}
R_{n}(x)=C_{n}(2\kappa _{2})^{-\left( \frac{1}{2}\widehat{D}\kappa _{2}-%
\frac{1}{2}-n\right) }\left( 2\kappa _{2}e^{-ix}\right) ^{\left( \frac{1}{2}%
\widehat{D}\kappa _{2}-\frac{1}{2}-n\right) }e^{-\kappa _{2}\exp
(-ix)}L_{n}^{2\left( \frac{1}{2}\widehat{D}\kappa _{2}-\frac{1}{2}-n\right)
}\left( 2\kappa _{2}e^{-ix}\right) ,
\end{equation}%
where $C_{n}$ is a normalizing factor.

\subsection{The Second Type of PT-Symmetric and Non-Hermitian Morse case}

We consider the strength parameters $V_{1}=\omega ^{2},$ $V_{2}=D,$ $V_{3}=0$
and $\alpha \rightarrow i\alpha $ where $\omega $ and $D$ are real
constants. The potential takes the form: 
\begin{equation}
V(x)=-\omega ^{2}e^{-2i\alpha x}+De^{-i\alpha x},
\end{equation}%
we get real spectrum for this kind of potentials:%
\begin{equation}
E_{n}=E_{0}\left[ \frac{1}{2}\frac{\sqrt{D}}{\omega }\kappa _{3}-n-\frac{1}{2%
}\right] ^{2},\text{ }n=0,1,2,\cdots ,n_{\max },
\end{equation}%
with%
\begin{equation}
\text{ }\kappa _{3}=\frac{r_{e}}{\hbar }\sqrt{2\mu D},\text{ }n_{\max }\leq 
\frac{1}{2}\left( \frac{r_{e}}{\hbar }\frac{\sqrt{D}}{\omega }\sqrt{2\mu D}%
-1\right) .
\end{equation}%
Also, this gives the wave functions as%
\begin{equation*}
R_{n}(x)=D_{n}(2\kappa _{3})^{-\left( \frac{1}{2}\frac{\sqrt{D}}{\omega }%
\kappa _{3}-\frac{1}{2}-n\right) }\left( 2\kappa _{3}e^{-i\alpha x}\right)
^{\left( \frac{1}{2}\frac{\sqrt{D}}{\omega }\kappa _{3}-\frac{1}{2}-n\right)
}e^{-\kappa _{3}\exp (-i\alpha x)}
\end{equation*}%
\begin{equation}
\times L_{n}^{2\left( \frac{1}{2}\frac{\sqrt{D}}{\omega }\kappa _{2}-\frac{1%
}{2}-n\right) }\left( 2\kappa _{3}e^{-i\alpha x}\right) ,
\end{equation}%
where $D_{n}$ is a normalizing factor.

\section{Results}

Firstly, in the constant mass limit, the numerically generated vibrational
bound state energies of the original Morse potential ($q=1,$ $V_{3}=0$ case)
for $H_{2},$ $LiH,$ $HCl$ and $CO$ molecules are found to be identical to
those given in Table 2 of Ref. [15] (available from the author upon
request). These numerical computations were performed using the model
parameters [6,10,13,14,16] shown in Table 2 with the order of the
eigenvalues represented by $n$ (the vibrational quantum number). It is found
that NU method, in the present study, can generate results similar to the
tridiagonal J-matrix representation [15] which is relatively cumbersome in
solving a matrix of dimension $N=100$ for $H_{2},$ $LiH$ and $HCl$ molecules
to even $N=200$ for $CO$ molecule. We have also calculated the total number
of bound states $n_{\max }=17,24,29$ and $83$ along with the whole
vibrational bound-state spectrum near the continuous zone for the above
molecules, respectively [56]. The bound state energy for the last state is
found to be $E_{n_{\max }}=-1.231\times 10^{-4},-1.303\times
10^{-3},-1.270\times 10^{-3}$ and $-5.533\times 10^{-7}eV$ for $H_{2},HCl,$ $%
LiH$ and $CO$ molecules, respectively. The relative accuracy can be of order 
$10^{-5}$ or less (up to five significant figures) for $LiH,$ $HCl$ and $CO$
molecules and $10^{-4}$-$10^{-5}$ for $H_{2}$ molecule (up to four-five
significant figures).

In addition, the known spectroscopic values in Table 2 are used to produce
the energy states for selected different arbitrary values of the vibrational 
$n$ and rotational $l$ angular momentum as shown in Table 3. We also list
analogous results obtained before by other methods (using the Pekeris
scheme) such as the NU [10], AIM [12], variational [6], SUSY [7] and EQR
[13,14] methods together with perturbative and variational quantum
mechanical methods like the tridiagonal J-matrix representation (TJM) [15],
the shifted $1/N$ expansion (SE) [9], the modified shifted $1/N$ expansion
(MSE) [9], the results of Duff and Rabitz (DR) [17], the hyper-virial
perturbation (HV) [8], the two-point quasi-rational approximation technique
(TQA) [16] and the Fourier grid Hamiltonian method (FGH) [18]. The quality
of the results is reassuring with the agreement between our numerical
results and those generated by other methods reaching up to four-five
significant digits. The current NU approximations to the ro-vibrational
energy bound-states are slightly improved from the previous NU
approximations probably for two simple reasons. The present calculations are
given to four significant digits and if they rounded off to
three-significant figures would coincide with those given before in Ref.
[10]. Throughout this numerical study, the parameter conversions in Table 1
into energy units ($eV$) might be the second reason. The conversions used
are $1$ $amu=931.502$ $MeV/c^{2},1cm^{-1}=1.23985\times 10^{-4}eV$ and $%
\hbar c=1973.29$ $eV\times $ $A^{\circ }$ (cf. pp. 791 in [57]). A first
look at Table 3 shows that the present approximations give results of
identical accuracy like the other well-known semi-analytic variational, EQR,
AIM and SUSY methods using the same Pekeris approximation for those fairly
small rotational quantum numbers $l.$ This is simply due to the Pekeris
approximation [5] where the centrifugal term is being approximated to
second-order in $r/r_{e}$ (i.e., at lower rotational energy, where $r\approx
r_{e}$). This is why the Pekeris approximation is valid only for very small
spatial variations from the inter-nuclear separation. The method loses its
accuracy for these higher ro-vibrational states. Nonetheless, our calculated
vibrational energy states are in higher agreement with the recently
calculated energy states (cf. Table 2 in Ref. [15]) for lower- and
higher-excitation energy states. Overmore, the ro-vibrational energy states
generated by non-perturbative NU, EQR, SUSY, AIM and variational methods are
nearly same and slightly different from those given by the cumbersome
perturbative and variational methods like HV, TJM, SE, MSE, TQA, FGH and DR,
which are not using Pekeris approximation, particularly for
higher-excitation ro-vibrational levels.

The further numerical calculations of the ro-vibrational energy states for
various quantum numbers $n$ and $l$ on the $CO,$ $LiH,$ $H_{2}$ and $HCl$
molecules [55] are also calculated (available from the author upon request)
using the generalized $q$-deformed Morse potential ($V_{3}=0$ case). The
variation of the deformation parameter $q$ in our model will produce much
more accurate information throughout these spectra about the structure and
dynamics of such diatomic molecules. For the case of multi-atomic molecules,
the generalization of Morse potentials can be expressed by the Empirical
Valence Bond (EVB) approach [45]. Connections of potential functions have
been extensively established for various combinations of pair potentials,
these have been largely confined to simple potentials such as the harmonic
[58], Leonard-Jones [59], Morse [1], Rydberg [60] and Buckingham [61]
potential functions (cf. Ref. [62]). For example, parametric connections
between the generalized Morse and Extended-Rydberg (ER) potential functions
have been attained. Since the number of parameters for ER exceeds those of
generalized Morse by 1, therefore only $V_{i}$ ($i=1,2$) are required for
converting the ER parameters into generalized Morse parameters but both the
Morse indices are needed to obtain $V_{1},$ $V_{2}$ and $V_{3}$ [45,63]$.$
The present potential functions have greater flexibility consisting of more
parameters. There are two sets of relationship between the Generalized Morse
and the ER parameters: (i) for the case where $V_{2}<0,$ and (ii) for the
case where $V_{2}>0.$ For example, the choice of low deformation [43] $q=-1$
provides $V_{2}<0,$ i.e., Morse-like potential, $V_{M}(r)=D_{e}\left(
e^{-2a(r-r_{e})}+2e^{-a(r-r_{e})}\right) $ with the position of a minimum
value approaching at $V_{M}(r_{e})=3D_{e}$ and strength ratio $\frac{V_{2}}{%
V_{1}}=-2.$ Whereas the high deformations $q=\pm 5$ provide the two
mentioned cases for the potential $V_{M}(r)=D_{e}\left( e^{-2a(r-r_{e})}\mp
10e^{-a(r-r_{e})}\right) $ of minima values approaching at $%
V_{M}(r_{e})=-9D_{e}$ and $11D_{e},$ strength ratios $\frac{V_{2}}{V_{1}}%
=\pm 10,$ respectively. Some other numerical energy calculations can be
estimated including the generalized $q$-deformed Morse potential ($V_{3}\neq
0$ case) [42] (available from the author upon the request). For example, the
choice of low deformations $q=\pm 1$ provide a solution for a Morse-like
potential, $V_{M}(r)=D_{e}\left( \pm 1-e^{-a(r-r_{e})}\right) ^{2}$ with a
minimum value approaching at $V_{M}(r_{e})=0$ and $4D_{e},$ respectively. By
this choice of parameters $V_{2}<0$ and $V_{2}>0$ for Be-S and H-Na
potentials, respectively, we can optimize the generalized Morse potential to
ER where good correlation is seen at and near equilibrium [45,62]. This wide
range of spectra for various deformed potential models might be necessary in
fitting the true experimental one. This approach could be also useful to
study further $Ar_{2},$ $O_{2},$ $I_{2}$ and $NO$ molecules and others.

\section{Final Remarks}

To summarize, we have used a parametric generalization of the NU method
derived for any exponential-type potential to obtain the bound state
solutions of the spatially-dependent mass Schr\"{o}dinger equation with any
rotational angular momentum quantum number $l$ for the generalized $q$%
-deformed Morse potential. In this paper, a suitable choice of a mass
function has also been proposed. The present calculations include energy
equation and the normalized wavefunctions expressed in terms of the Jacobi
and Laguerre polynomials. We find a general equation for the ro-vibrational
bound-state energy eigenvalues true for the currently proposed mass function
given in (22) and in terms of three potential strength parameters $V_{i}$
(where $i=1,2,3$). Hence, with selected values of the parameter $\delta ,$
we can obtain a family of solutions. The non-relativistic limit of the
solution is being obtained by an appropriate choice of the parameter $\delta
\rightarrow 0$ in the mass function. Obviously, we can generate
non-relativistic bound state solutions for the rotating Morse potential when
the deformation parameters $q=1.$ The numerical application of the method to
diatomic molecules demonstrates that the values obtained are in high
agreement with other methods and numerical data at low-rotational excitation
energy. This provides an alternative systematic procedure to calculate the
energy eigenvalues with a reasonable accuracy and also considered as a
suitable method in the treatment of such potentials with a varying mass
functions as well. It is worth noting that the analytical results presented
here allow one calculating the energy eigenvalues as well as wavefunctions
in a very simple way, with very high accuracy for lower-excitation
rotational levels and for lower- and higher-excitation vibrational levels,
probably enough for most of the applications known until now. The real
advantages of our semi-analytic method are systematic, highly accurate,
handy, easily implemented and not cumbersome as most of the other well-known
methods mentioned before in this paper. We believe that the procedures could
be easily extended to other short-range as well as long-range potentials.

\acknowledgments The author thanks the anonymous kind referees for the
constructive comments and suggestions that have improved the paper greatly.
He is also grateful for the partial support provided by the Scientific and
Technological Research Council of Turkey (T\"{U}B\.{I}TAK).

\appendix

\section{Parametric Generalization of the NU Method}

We complement the theoretical formulation of the NU method in presenting the
essential polynomials, energy equation and wave functions together with
their relevant constants as follows.

(i) The key polynomials:%
\begin{equation}
\pi (z)=c_{4}+c_{5}z-\left[ \left( \sqrt{c_{9}}+c_{3}\sqrt{c_{8}}\right) z-%
\sqrt{c_{8}}\right] ,
\end{equation}%
\begin{equation}
k=-\left( c_{7}+2c_{3}c_{8}\right) -2\sqrt{c_{8}c_{9}}.
\end{equation}%
\begin{equation}
\tau (z)=1-\left( c_{2}-2c_{5}\right) z-2\left[ \left( \sqrt{c_{9}}+c_{3}%
\sqrt{c_{8}}\right) z-\sqrt{c_{8}}\right] ,
\end{equation}%
\begin{equation}
\tau ^{\prime }(z)=-2c_{3}-2\left( \sqrt{c_{9}}+c_{3}\sqrt{c_{8}}\right) <0,
\end{equation}%
(ii) The energy equation:%
\begin{equation}
\left( c_{2}-c_{3}\right) n+c_{3}n^{2}-\left( 2n+1\right) c_{5}+\left(
2n+1\right) \left( \sqrt{c_{9}}+c_{3}\sqrt{c_{8}}\right) +c_{7}+2c_{3}c_{8}+2%
\sqrt{c_{8}c_{9}}=0.
\end{equation}%
(iii) The wave functions:%
\begin{equation}
\rho (z)=z^{c_{10}}(1-c_{3}z)^{c_{11}},
\end{equation}%
\begin{equation}
\phi (z)=z^{c_{12}}(1-c_{3}z)^{c_{13}},
\end{equation}%
\begin{equation}
y_{n}(z)=P_{n}^{\left( c_{10},c_{11}\right) }(1-2c_{3}z),
\end{equation}%
\begin{equation}
u(z)=\mathcal{N}_{n}z^{c_{12}}(1-c_{3}z)^{c_{13}}P_{n}^{\left(
c_{10},c_{11}\right) }(1-2c_{3}z),
\end{equation}%
where $P_{n}^{\left( \alpha ,\beta \right) }(1-2s)$ are the Jacobi
polynomials with%
\begin{equation}
P_{n}^{\left( \alpha ,\beta \right) }(1-2s)=\frac{\left( \alpha +1\right)
_{n}}{n!}%
\begin{array}{c}
_{2}F_{1}%
\end{array}%
\left( -n,1+\alpha +\beta +n;\alpha +1;s\right) ,
\end{equation}%
and $\mathcal{N}_{n}$ is a normalizing factor. Also, the above wavefunctions
can be expressed in terms of the hypergeometric function as%
\begin{equation}
u(z)=\mathcal{N}_{n}z^{c_{12}}(1-c_{3}z)^{c_{13}}%
\begin{array}{c}
_{2}F_{1}%
\end{array}%
\left( -n,1+c_{10}+c_{11}+n;c_{10}+1;c_{3}z\right) ,
\end{equation}%
where $%
\begin{array}{c}
_{2}F_{1}%
\end{array}%
\left( a,b;c;z\right) =\frac{\Gamma (c)}{\Gamma (a)\Gamma (b)}%
\dsum\limits_{p=0}^{\infty }\frac{\Gamma (a+p)\Gamma (b+p)}{\Gamma (c+p)}%
\frac{z^{p}}{p!}.$ We can determine the normalization constant using the
condition $\dint\limits_{0}^{\infty }u^{2}(r)dr=1$ and [64]%
\begin{equation}
\dint\limits_{0}^{1}\left( 1-s\right) ^{\mu -1}s^{\nu -1}%
\begin{array}{c}
_{2}F_{1}%
\end{array}%
\left( \alpha ,\beta ;\gamma ;as\right) dz=\frac{\Gamma (\mu )\Gamma (\nu )}{%
\Gamma (\mu +\nu )}%
\begin{array}{c}
_{3}F_{2}%
\end{array}%
\left( \nu ,\alpha ,\beta ;\mu +\nu ;\gamma ;a\right) .
\end{equation}

(iv) The relevant constants:%
\begin{equation*}
c_{4}=\frac{1}{2}\left( 1-c_{1}\right) ,\text{ }c_{5}=\frac{1}{2}\left(
c_{2}-2c_{3}\right) ,\text{ }c_{6}=c_{5}^{2}+A,
\end{equation*}%
\begin{equation*}
\text{ }c_{7}=2c_{4}c_{5}-B,\text{ }c_{8}=c_{4}^{2}+C,\text{ }%
c_{9}=c_{3}\left( c_{7}+c_{3}c_{8}\right) +c_{6},
\end{equation*}%
\begin{equation*}
c_{10}=c_{1}+2c_{4}+2\sqrt{c_{8}}-1,\text{ }c_{11}=1-c_{1}-2c_{4}+\frac{2}{%
c_{3}}\sqrt{c_{9}},
\end{equation*}%
\begin{equation}
c_{12}=c_{4}+\sqrt{c_{8}},\text{ }c_{13}=-c_{4}+\frac{1}{c_{3}}\left( \sqrt{%
c_{9}}-c_{5}\right) .
\end{equation}%
$\label{appendix}$

\section{The Pekeris Approximation for the Rotational Morse Potential}

In this appendix, we present the Pekeris approximation [5,10] performed up
to the second order term in $r/r_{e}$. We get started by making the change
of parameters and coordinates as follows: $x=\left( r-r_{e}\right) /r_{e}$
and expanding around $x=0$ ($r=r_{e}$) to obtain:%
\begin{equation}
V_{rot}(x)=\frac{\gamma }{(1+x)^{2}}=\gamma _{1}\left[ 1-2x+3x^{2}+O(x^{3})%
\right] ,\text{ }\gamma =\frac{l(l+1)}{r_{e}^{2}},
\end{equation}%
where the first few terms are sufficient for the lower-excitation rotational
states since $r\approx r_{e}.$ The corresponding rotational term expressed
in the exponential form up to the second order is 
\begin{equation}
\widetilde{V}_{rot}(x)=\gamma \left( a_{0}+a_{1}e^{-\alpha
z}+a_{2}e^{-2\alpha z}\right) ,
\end{equation}%
\begin{equation*}
\widetilde{V}_{rot}(x)=\gamma \left[ a_{0}+a_{1}\left( 1-\alpha x+\frac{%
\left( \alpha x\right) ^{2}}{2!}-O(x^{3})\right) +a_{2}\left( 1-2\alpha x+%
\frac{\left( 2\alpha x\right) ^{2}}{2!}-O(x^{3})\right) \right] ,
\end{equation*}%
\begin{equation}
=\gamma \left[ \sum_{i=0}^{2}a_{i}-\alpha \left( a_{1}+2a_{2}\right)
x+\alpha ^{2}\left( \frac{a_{1}}{2}+2a_{2}\right) x^{2}-O(x^{3})\right] ,
\end{equation}%
where $a_{i}$ ($i=1,2,3$) are the expansion coefficients. Comparing (B.1)
with (B.4), we obtain specific values for the expansion coefficients as%
\begin{equation}
a_{0}=1-\frac{3}{\alpha }\left( 1-\frac{1}{\alpha }\right) ,\text{ }a_{1}=%
\frac{2}{\alpha }\left( 2-\frac{3}{\alpha }\right) ,\text{ }a_{2}=-\frac{1}{%
\alpha }\left( 1-\frac{3}{\alpha }\right) ,\text{ }\alpha =ar_{e}.
\end{equation}%
On the other hand, we repeat similar procedures to obtain an exponential
expansion for the term $1/r$ around $x=0$ ($r=r_{e}$) as follows: 
\begin{equation}
\frac{1}{r}=\frac{1}{r_{e}(1+x)}=\frac{1}{r_{e}}\left[ 1-x+x^{2}-O(x^{3})%
\right] ,
\end{equation}%
or equivalently in the exponential expansion:%
\begin{equation*}
\frac{1}{r}=\frac{1}{r_{e}}\left( b_{0}+b_{1}e^{-\alpha z}+b_{2}e^{-2\alpha
z}\right) ,
\end{equation*}%
\begin{equation}
=\frac{1}{r_{e}}\left[ \sum_{i=0}^{2}b_{i}-\alpha \left( b_{1}+2b_{2}\right)
x+\alpha ^{2}\left( \frac{b_{1}}{2}+2b_{2}\right) x^{2}-O(x^{3})\right] ,
\end{equation}%
Thus, comparing (B.5) with (B.6), we obtain the following expansion
coefficients as%
\begin{equation}
b_{0}=1-\frac{1}{\alpha }\left( \frac{3}{2}-\frac{1}{\alpha }\right) ,\text{ 
}b_{1}=\frac{2}{\alpha }\left( 1-\frac{1}{\alpha }\right) ,\text{ }b_{2}=-%
\frac{1}{\alpha }\left( \frac{1}{2}-\frac{1}{\alpha }\right) ,\text{ }\alpha
=ar_{e}.
\end{equation}

\newpage\ 

{\normalsize 
}

\bigskip

\baselineskip= 2\baselineskip
\bigskip \newpage

\bigskip {\normalsize 
}

\baselineskip= 2\baselineskip
\bigskip \newpage 
\begin{table}[tbp]
\caption{The specific values for the parametric constants necessary for the
present potential.}%
\begin{tabular}{llll}
\tableline Constant & Value & Constant & Value \\ 
\tableline$c_{1}$ & 1 & $c_{2}$ & $\delta $ \\ 
$c_{3}$ & $\delta $ & c$_{4}$ & $0$ \\ 
$c_{5}$ & $-\frac{1}{2}\delta $ & $c_{6}$ & $\frac{1}{4}\left( \delta
^{2}+4\beta _{1}\right) $ \\ 
$c_{7}$ & $-\beta _{2}$ & $c_{8}$ & $\varepsilon _{nl}^{2}$ \\ 
$c_{9}$ & $\frac{1}{4}\delta ^{2}\xi ^{2}$ & $c_{10}$ & $2\varepsilon _{nl}$
\\ 
$c_{11}$ & $\xi $ & $c_{12}$ & $\varepsilon _{nl}$ \\ 
$c_{13}$ & $\frac{1}{2}\left( 1+\xi \right) $ & $A$ & $\beta _{1}$ \\ 
$B$ & $\beta _{2}$ & $C$ & $\varepsilon _{nl}^{2}$ \\ 
\tableline &  &  &  \\ 
&  &  & 
\end{tabular}%
\end{table}
\ 
\begin{table}[tbp]
\caption{Model parameters for $CO,$ $LiH,$ $H_{2}$ and $HCl$ diatomic
molecules in our study as obtained from the cited sources. }%
\begin{tabular}{lllll}
\tableline Parameters & $CO$ [13] & $LiH$ [13] & $H_{2}$ [16] & $HCl$ [13]
\\ 
\tableline$D_{0}$ $(cm^{-1})$ & $90540$ & $20287$ & $38266$ & $37255$ \\ 
$a$ $(A^{\circ })^{-1}$ & $2.2994$ & $1.1280$ & $1.9426$ & $1.8677$ \\ 
$r_{0}$ $(A^{\circ })$ & $1.1283$ & $1.5956$ & $0.7416$ & $1.2746$ \\ 
$\mu $ (amu) & $6.8606719$ & $0.8801221$ & $0.50391$ & $0.9801045$ \\ 
\tableline &  &  &  & 
\end{tabular}%
\end{table}

\bigskip 
\begin{table}[tbp]
\caption{Bound-state energy eigenvalues ($-E_{nl}$) for the $H_{2},$ $LiH,CO$
and $HCl$ molecules (in $eV$) for different values of $\ $the vibrational $n$
and rotational $l$ quantum numbers in the usual Morse potential ($q=1,$ $%
V_{3}=0$).}%
\begin{tabular}{llllllllllll}
\tableline$n$ & $l$ & This work & SUSY [7] & AIM [12] & HV [8] & DR [17] & 
MSE [9] & TJM [15] & SE [9] & V [6] & TQA [16] \\ 
\tableline &  &  &  &  &  & $H_{2}$ &  &  &  &  &  \\ 
\tableline$0$ & $0$ & $4.47601$ & $4.47601$ & $4.47601$ & $4.47601$ & $%
4.4762 $ & $4.4760$ & $4.4760131$ & $4.4749$ & $4.4758$ & $4.4760084$ \\ 
& $5$ & $4.25880$ & $4.25880$ & $4.25880$ & $4.25901$ & $4.2592$ & $4.2590$
& $4.2590180$ & $4.2589$ & $4.2563$ & $4.2590038$ \\ 
& $10$ & $3.72194$ & $3.72193$ & $3.72193$ & $3.72473$ & $3.7251$ & $3.7247$
& $3.7247471$ & $3.7247$ & $3.7187$ & $3.7247181$ \\ 
$5$ & $0$ & $2.22052$ & $2.22051$ & $2.22052$ & $2.22051$ & $2.218$ & $%
2.2205 $ & $2.2205369$ &  &  &  \\ 
& $5$ & $2.04355$ & $2.04353$ & $2.04355$ & $2.05285$ & $2.050$ & $2.0430$ & 
$2.0528808$ &  &  &  \\ 
& $10$ & $1.60391$ & $1.60389$ & $1.60391$ & $1.65265$ & $1.650$ & $1.6535$
& $1.6526902$ &  &  &  \\ 
$7$ & $0$ & $1.53744$ & $1.53743$ & $1.53744$ & $1.53743$ &  & $1.5374$ & $%
1.5374552$ &  &  &  \\ 
& $5$ & $1.37656$ & $1.37654$ & $1.37656$ & $1.39263$ &  & $1.3932$ & $%
1.3926614$ &  &  &  \\ 
& $10$ & $0.97581$ & $0.97578$ & $0.97581$ & $1.05265$ &  & $1.0552$ & $%
1.0526836$ &  &  &  \\ 
\tableline$n$ & $l$ & This work & NU [10] & EQR [13] & SUSY [7] & AIM [12] & 
MSE [9] & TJM [15] & FGH [18] & SE [9] & V [6] \\ 
\tableline &  &  &  &  &  & $LiH$ &  &  &  &  &  \\ 
\tableline$0$ & $0$ & $2.42886$ & $2.4287$ & $2.42886$ & $2.42886$ & $2.4289$
& $2.4280$ & $2.4288627$ & $2.42886$ & $2.4278$ & $2.4291$ \\ 
& $5$ & $2.40133$ & $2.4012$ & $2.40133$ & $2.40133$ & $2.4013$ & $2.4000$ & 
$2.4013352$ & $2.40133$ & $2.3999$ & $2.4014$ \\ 
& $10$ & $2.32884$ & $2.3287$ & $2.32883$ & $2.32883$ & $2.3288$ & $2.3261$
& $2.3288530$ & $2.32885$ & $2.3261$ & $2.3287$ \\ 
$5$ & $0$ & $1.64771$ & $1.6476$ & $1.64772$ & $1.64772$ & $1.6477$ & $%
1.6402 $ & $1.6477149$ & $1.64772$ & $1.6242$ &  \\ 
& $5$ & $1.62377$ & $1.6236$ & $1.62377$ & $1.62377$ & $1.6238$ & $1.6160$ & 
$1.6239497$ & $1.62395$ & $1.6074$ &  \\ 
& $10$ & $1.56074$ & $1.5606$ & $1.56074$ & $1.56074$ & $1.5607$ & $1.5525$
& $1.5615114$ & $1.56152$ & $1.5479$ &  \\ 
$7$ & $0$ & $1.37756$ & $1.3774$ & $1.37757$ & $1.37756$ & $1.3776$ & $%
1.3862 $ & $1.3775588$ & $1.37756$ & $1.3424$ &  \\ 
& $5$ & $1.35505$ & $1.3549$ & $1.35505$ & $1.35505$ & $1.3550$ & $1.3456$ & 
$1.3553770$ & $1.35538$ & $1.3309$ &  \\ 
& $10$ & $1.29580$ & $1.2956$ & $1.29581$ & $1.29580$ & $1.2958$ & $1.2865$
& $1.2971612$ & $1.29715$ & $1.2781$ &  \\ 
\tableline &  &  &  &  &  &  &  &  &  &  & 
\end{tabular}%
\end{table}

\bigskip 
\begin{table}[tbp]
\caption{Continue.}%
\begin{tabular}{llllllllllll}
\tableline$n$ & $l$ & This work & SUSY [7] & AIM [12] & NU [10] & EQR [13] & 
MSE [9] & TJM [15] & SE [9] & FGH [18] & V [6] \\ 
\tableline &  &  &  &  &  & $CO$ &  &  &  &  &  \\ 
\tableline$0$ & $0$ & $11.0915$ & $11.0915$ & $11.0915$ & $11.091$ & $%
11.0915 $ & $11.092$ & $11.0915353$ & $11.091$ & $11.0915$ & $11.093$ \\ 
& $5$ & $11.0844$ & $11.0844$ & $11.0845$ & $11.084$ & $11.0844$ & $11.084$
& $11.0843875$ & $11.084$ & $11.0844$ & $11.084$ \\ 
& $10$ & $11.0653$ & $11.0653$ & $11.0653$ & $11.065$ & $11.0653$ & $11.065$
& $11.0653334$ & $11.065$ & $11.0653$ & $11.0653$ \\ 
$5$ & $0$ & $9.79518$ & $9.79519$ & $9.7952$ & $9.795$ & $9.79519$ & $9.795$
& $9.7951838$ & $9.788$ & $9.79519$ &  \\ 
& $5$ & $9.78833$ & $9.78834$ & $9.7883$ & $9.788$ & $9.78835$ & $9.788$ & $%
9.7883443$ & $9.782$ & $9.78835$ &  \\ 
& $10$ & $9.77009$ & $9.77010$ & $9.7701$ & $9.769$ & $9.77011$ & $9.770$ & $%
9.7701124$ & $9.765$ & $9.77011$ &  \\ 
$7$ & $0$ & $9.29918$ & $9.29920$ & $9.2992$ & $9.299$ & $9.29920$ & $9.299$
& $9.2991935$ & $9.286$ & $9.29920$ &  \\ 
& $5$ & $9.29246$ & $9.29248$ & $9.2925$ & $9.292$ & $9.29248$ & $9.292$ & $%
9.2924786$ & $9.281$ & $9.29248$ &  \\ 
& $10$ & $9.27455$ & $9.27458$ & $9.2745$ & $9.274$ & $9.27457$ & $9.274$ & $%
9.2745791$ & $9.265$ & $9.27458$ &  \\ 
\tableline$n$ & $l$ & This work & V [6] & EQR [13] & SUSY [7] & AIM [12] & 
MSE [9] & TJM [15] & FGH [18] & SE [9] &  \\ 
\tableline &  &  &  &  &  & $HCl$ &  &  &  &  &  \\ 
\tableline$0$ & $0$ & $4.43556$ & $4.4360$ & $4.43556$ & $4.43556$ & $4.4356$
& $4.4355$ & $4.4355522$ & $4.43556$ & $4.4352$ &  \\ 
& $5$ & $4.39682$ & $4.3971$ & $4.39681$ & $4.39681$ & $4.3968$ & $4.3968$ & 
$4.3968066$ & $4.39682$ & $4.3967$ &  \\ 
& $10$ & $4.29408$ & $4.2940$ & $4.28407$ & $4.28408$ & $4.2841$ & $4.2940$
& $4.2940628$ & $4.28409$ & $4.2939$ &  \\ 
$5$ & $0$ & $2.80506$ &  & $2.80507$ & $2.80508$ & $2.8051$ & $2.8046$ & $%
2.8049687$ & $2.80508$ &  &  \\ 
& $5$ & $2.77209$ &  & $2.77210$ & $2.77211$ & $2.7721$ & $2.7718$ & $%
2.7721880$ & $2.77230$ &  &  \\ 
& $10$ & $2.68471$ &  & $2.68472$ & $2.68473$ & $2.6847$ & $2.6850$ & $%
2.6853673$ & $2.68549$ &  &  \\ 
$7$ & $0$ & $2.25701$ &  & $2.25702$ & $2.25703$ & $2.2570$ & $2.2565$ & $%
2.2568924$ & $2.25703$ &  &  \\ 
& $5$ & $2.22634$ &  & $2.22636$ & $2.22636$ & $2.2263$ & $2.2262$ & $%
2.2265969$ & $2.22673$ &  &  \\ 
& $10$ & $2.14511$ &  & $2.14512$ & $2.14513$ & $2.1451$ & $2.1461$ & $%
2.1464148$ & $2.14656$ &  &  \\ 
\tableline &  &  &  &  &  &  &  &  &  &  & 
\end{tabular}%
\end{table}

\end{document}